\documentclass[iop]{emulateapj}
\usepackage{graphicx}

\begin{document}

\title{UV anti-reflection coatings for use in silicon detector design}
\author{Erika~T.~Hamden$^1$, Frank~Greer$^2$, Michael~E.~Hoenk$^2$, Jordana~Blacksberg$^2$, Matthew~R.~Dickie$^2$, Shouleh~Nikzad$^2$, Christopher~D.~Martin$^3$, David~Schiminovich$^1$}

\affil{$^1$Department of Astronomy, Columbia University,  550 W 120th St, New York, NY 10025, USA,  hamden@astro.columbia.edu}
\affil{$^2$Jet Propulsion Laboratory, California Institute of Technology,  91109}
\affil{$^3$Department of Astronomy, California Institute of Technology, 1200 East California Blvd, Pasadena, CA 91125, USA}

\begin{abstract}

We report on the development of coatings for a CCD detector optimized for use in a fixed dispersion UV spectrograph.  Due to the rapidly changing index of refraction of Si, single layer broadband anti-reflection coatings are not suitable to increase quantum efficiency at all wavelengths of interest.  Instead, we describe a creative solution that provides excellent performance over UV wavelengths.  We describe progress in the development of a CCD detector with theoretical quantum efficiencies (QE) of greater than 60\% at wavelengths from 120 to 300nm.  This high efficiency may be reached by coating a backside illuminated, thinned, delta-doped CCD with a series of thin film anti-reflection coatings. The materials tested include MgF$_2$ (optimized for highest performance from 120-150nm), SiO$_2$ (150-180nm), Al$_2$O$_3$(180-240nm), MgO (200-250nm), and HfO$_2$ (240-300nm).  A variety of deposition techniques were tested and a selection of coatings which minimized reflectance on a Si test wafer were applied to live devices.  We also discuss future uses and improvements, including graded and multi-layer coatings.

\end{abstract}

\section{Introduction}

Charged Coupled Devices (CCDs) were first invented at Bell Labs in 1969, and have since revolutionized imaging.  The CCD's ability to quickly and efficiently digitize data, its relatively low noise capabilities, and a sensitivity 100 times that of film meant that they quickly became indispensable to modern astronomy \citep{2001J}.  For UV astronomy, CCDs have not historically been a success.  While film is sensitive at nearly all wavelengths of light, unmodified CCDs present a number of deficiencies.  The front circuitry of a CCD is absorptive at UV wavelengths.  For high efficiency, CCDs are back illuminated, however the Si substrate is in turn highly reflective.  More crucially, UV photons have a very short absorption depth in Si.  The resulting electron-hole pairs find traps at the surface and never reach the gates for eventual readout.  These characteristics have created problems in developing efficient CCD based UV detectors.  One notable work-around to these issues was used for the Wide Field Planetary Camera 2 (WFPC2) on the Hubble Space Telescope (HST) \citep{1995H}.  WFPC2 used thick, front illuminated CCDs but coated them in a layer of the UV phosphor Lumogen, which provided a UV response of 10-15\% from 200 to 400nm by down converting the UV photons to 510-580nm photons.  The current camera on HST, WFC3, also utilizes a CCD (thinned and backside illuminated with a charged backside) and AR-coating for near UV, but suffers somewhat from quantum efficiency (QE) hysteresis (\citep{2010B}, \citep{2009C}).  Other types of UV detectors have been developed and are in use on current missions, including micro-channel plates (MCPs) (such as JUNO \citep{2008Gladstone}, FUSE \citep{2000Moos}, GALEX \citep{2003Martin}, ALICE \citep{2008Stern}, and FIREBall \citep{2008Tuttle}, to name a few).  While their photon-counting capabilities make them useful and at times the only option, MCPs still suffer from low quantum efficiency (QE from 25\% in the FUV (1344-1786 \AA) down to 8\% in NUV (1771-2831 \AA), for the GALEX MCPs) and are challenging to produce and utilize.  In this work, we examine an improved CCD which is able to overcome the difficulties described above.

Improvements to traditional CCD design beyond optical use have been made in recent years with the advent of delta-doped backside illuminated CCDs \citep{1992Hoenk}.  Using a custom backside treatment developed at JPL, these modified CCDs achieve near 100\% internal quantum efficiency from the ultraviolet to the near infrared.  This is achieved by thinning the backside surface to the epilayer (which can vary between 5-20$\mu$m depending on CCD type) and then delta-doping the thinned backside layer.  Delta-doping passivates the silicon surface and produces a highly stable, QE-pinned detector.  The atomic-scale precision of MBE growth is used to embed a layer of dopant atoms with the silicon lattice, which, when ionized, forms a sheet of negative charge that isolates the surface from the builk.  The boron atoms used in this case have an areal density of 2 $\times 10^{14}$ cm$^{-2}$, and are confined by MBE growth to nearly a single atomic plane located 1.5nm below the silicon surface.  These adaptations have greatly expanded the spectral range of modified CCDs, while also dramatically increasing QE and stability compared to their ready-made brethren \citep{1994Nikzad}.  Further work on anti-reflection (AR) coatings in the visible and near-infrared has been successful (\citep{2006B}, \citep{1994Nikzad}), prompting the current efforts to create similar AR coatings for the ultraviolet.  The technical limitations preventing new, innovative designs in UV detectors are quickly receding, and further development in the manufacturability of these thinned and doped devices is underway \citep{2009Hoenk}.  This paper describes work on the design, modeling, materials considerations and sample measurements on live devices as part of a larger effort for development of high QE, high performance far UV detectors.

This paper focuses on the development of an astronomical detector optimized for use with a monolithic UV spectrograph from 120 to 300nm.  Due to the highly variable indices of Si, no single material or thickness is sufficient to achieve high QE over the whole range.  The lack of a suitable broadband AR-coating in the UV does not rule out alternatives, and one can still achieve excellent performance through a creative use of coatings.  Sacrificing a single coating, and thus broadband imaging potential, and instead using a series of sequential narrowband coatings allows for higher QE overall.  Following this idea, a delta-doped CCD is to be coated in sections using different materials (an alternative would be a mosaic of devices, each with one coating).  This tiling lends itself most readily to a fixed spectrograph, where only one wavelength of light hits a particular region on the CCD.  Spectrographs with a fixed grating are becoming more common since they are relatively easy and low cost to mass-produce.  Space based instruments may also employ a fixed grating to reduce complexity and cost.

We have selected MgF$_2$, MgO, HfO$_2$, Al$_2$O$_3$, and SiO$_2$ to test as suitable AR coatings.  The materials chosen reflect the unique requirements of a UV AR coating, including a favorable index of refraction and low absorption in the desired waveband.  One must be able to deposit the film in a uniform way, while not causing damage to the CCD itself.  This eliminates electron beam evaporation (a common choice for dielectric coatings) as a potential technique because it causes X-ray damage to the CCDs.  We have tested sputtering, atomic layer deposition (ALD), and thermal evaporation techniques for consistency and measured the reflectance of films on a substrate of bare Si.  This is a low cost, faster alternative to testing on live devices.  The downside is that testing is limited to reflectance off this surface, which necessarily omits any absorption losses.  We then compare our measurements to the theoretical models.  A further discussion of deposition techniques and their effect on film quality will be presented in a forthcoming paper (Greer, 2011 (submitted)).

We also quickly report on the application of films onto live devices.  A detailed account of these results are under prepartion to be submitted for publication (Nikzad, 2011 (submitted)).  Briefly, thinned and delta-doped standard and EMCCDs were both used in a variety of tests to measure absolute QE.  EMCCDs are an advancement in CCD techonology which enables photon-counting.  A longer explanation is provided in Section \ref{sec:devices}.  These live device tests provide a more realistic view of the performance of these films than simple reflectance tests.  With live devices we are able to directly measure the effect the AR coating has on transmission into the the Si.  Papers are in preparation on the stability and testing of these devices, as well as on detailed live results and future FUV detector technology (Hoenk, 2011 (submitted); Nikzad 2011 (submitted); \citep{2011Jacquot}).

Here we report on the growth of these films and the optical testing of reflectance which we have performed.  In Section \ref{sec:Techniques} we discuss how films were deposited and what materials were chosen.  In Section \ref{sec:testing} we discuss the conditions for reflectance testing, following with a discussion of our results in Section \ref{sec:results}.  We also include a discussion of the data from live devices in Section \ref{sec:devices}.  Finally we briefly look to future work and more advanced coating models in Section \ref{sec:future}.

\section{Techniques}
\label{sec:Techniques}

\begin{figure}
\centering
\includegraphics[width=0.45\textwidth,bb= 56 231 576 576]{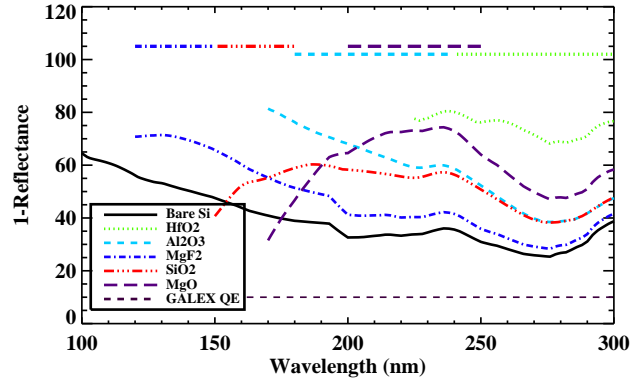}
\caption{Theoretical 1-Reflectance of all films.  Colored bars indicate the wavelength range over which each material is being considered.  The dashed line at the bottom represents the average QE of the GALEX micro-channel plate detector.}\label{fig:ideal}
\end{figure}

All film depositions were performed at the Jet Propulsion Laboratory (JPL) using thermal evaporation, Atomic Layer Deposition (ALD), and Radio Frequency (RF) dielectric sputtering.  We have used thermal evaporation to deposit layers of MgF$_2$ and RF sputtering to deposit layers of MgO and SiO$_2$.  RF sputtering and ALD were both used for making films of HfO$_2$, and Al$_2$O$_3$.  All films were grown on 1''  $\langle$100$\rangle$ P/B 1-20 Ohm-cm single-side polished wafers of Si, as a proxy for the actual device.  The thickness of each material has been selected to minimize reflectance in a specific wavelength range.  Calculations of predicted reflectance were done using the TFcalc$^{TM}$ software package, with a bare Si substrate and are shown in Figure \ref{fig:ideal}, along with the current average QE of the GALEX UV space telescope.  

\begin{figure}
\centering
\includegraphics[width=0.45\textwidth,bb= 56 231 576 576]{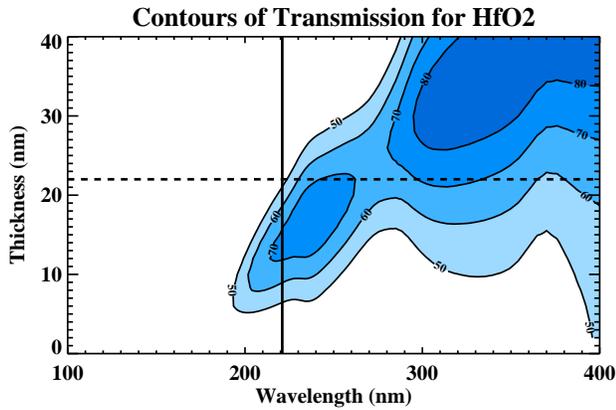}
\caption{Theoretical Transmission for a variety of thicknesses of HfO$_2$.  Contour lines begin at 50 percent transmission and increase in increments of 10 percent.  Horizontal line indicates thickness target.  Vertical line indicates absorption edge.  Absorption increases rapidly as wavelength decreases.}\label{fig:peakHfO2}
\end{figure}

\begin{figure}
\centering
\includegraphics[width=0.45\textwidth,bb= 56 231 576 576]{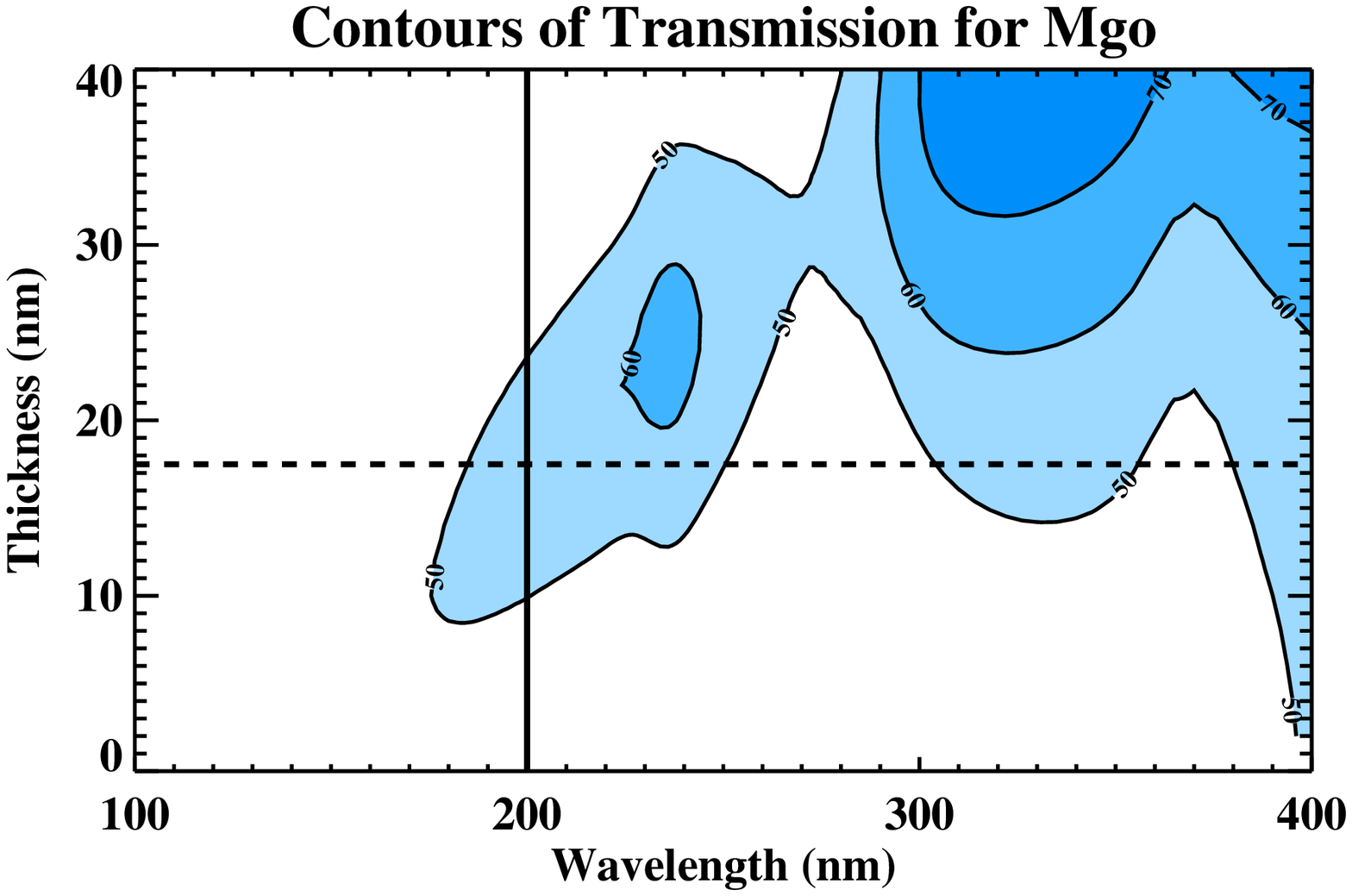}
\caption{Theoretical Transmission for a variety of thicknesses of MgO.  Contour lines begin at 50 percent transmission and increase in increments of 10 percent.  Horizontal line indicates thickness target.  Vertical line indicates absorptionedge.  Absorption increases rapidly as wavelength decreases.}\label{fig:peakMgo}
\end{figure}

\begin{figure}
\centering
\includegraphics[width=0.45\textwidth,bb= 56 231 576 576]{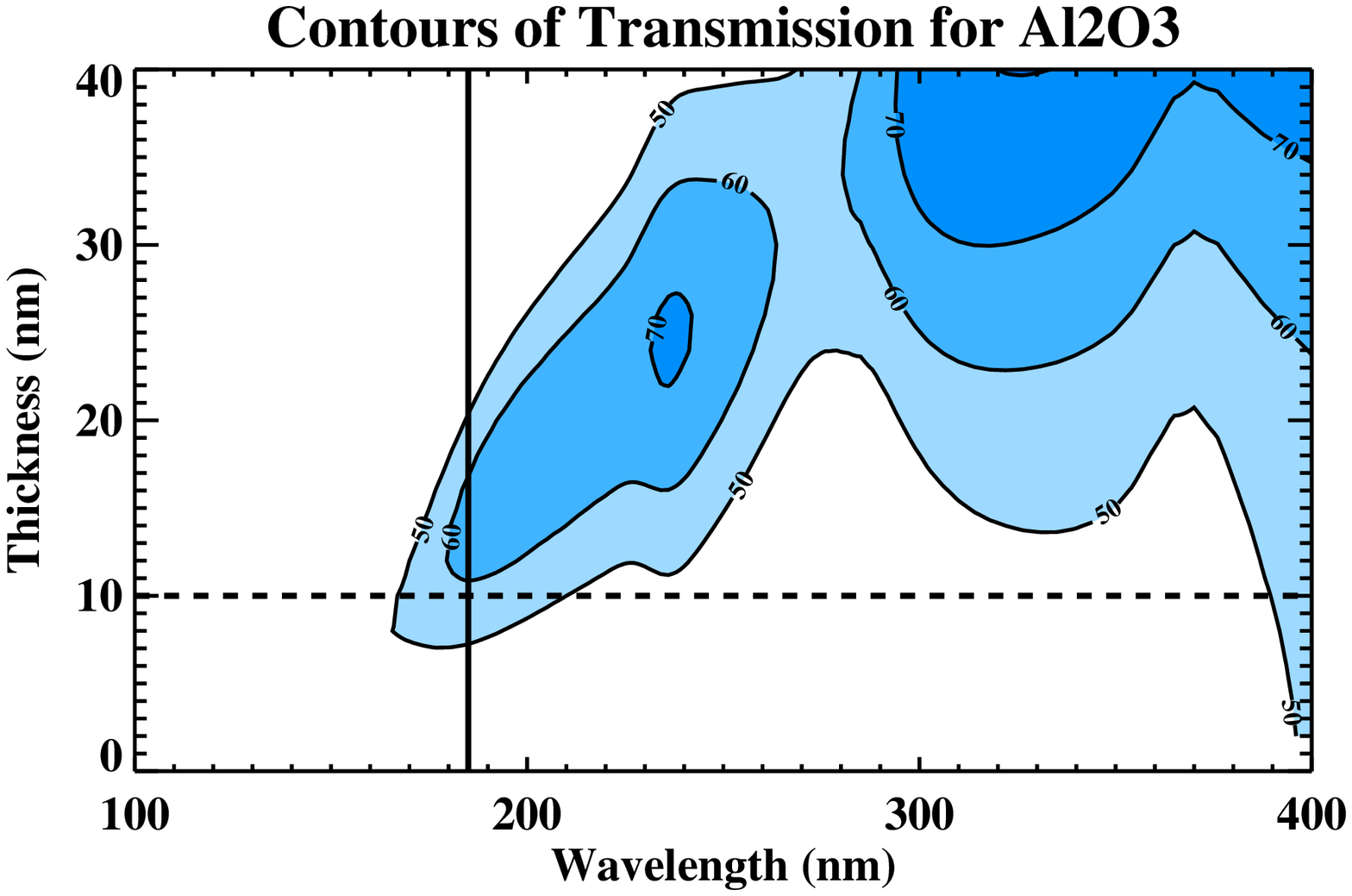}
\caption{Theoretical Transmission for a variety of thicknesses of Al$_2$O$_3$.  Contour lines begin at 50 percent transmission and increase in increments of 10 percent.  Horizontal line indicates thickness target.  Vertical line indicates absorption edge.  Absorption increases rapidly as wavelength decreases.}\label{fig:peakAl2O3}
\end{figure}

\begin{figure}
\centering
\includegraphics[width=0.45\textwidth,bb= 56 231 576 576]{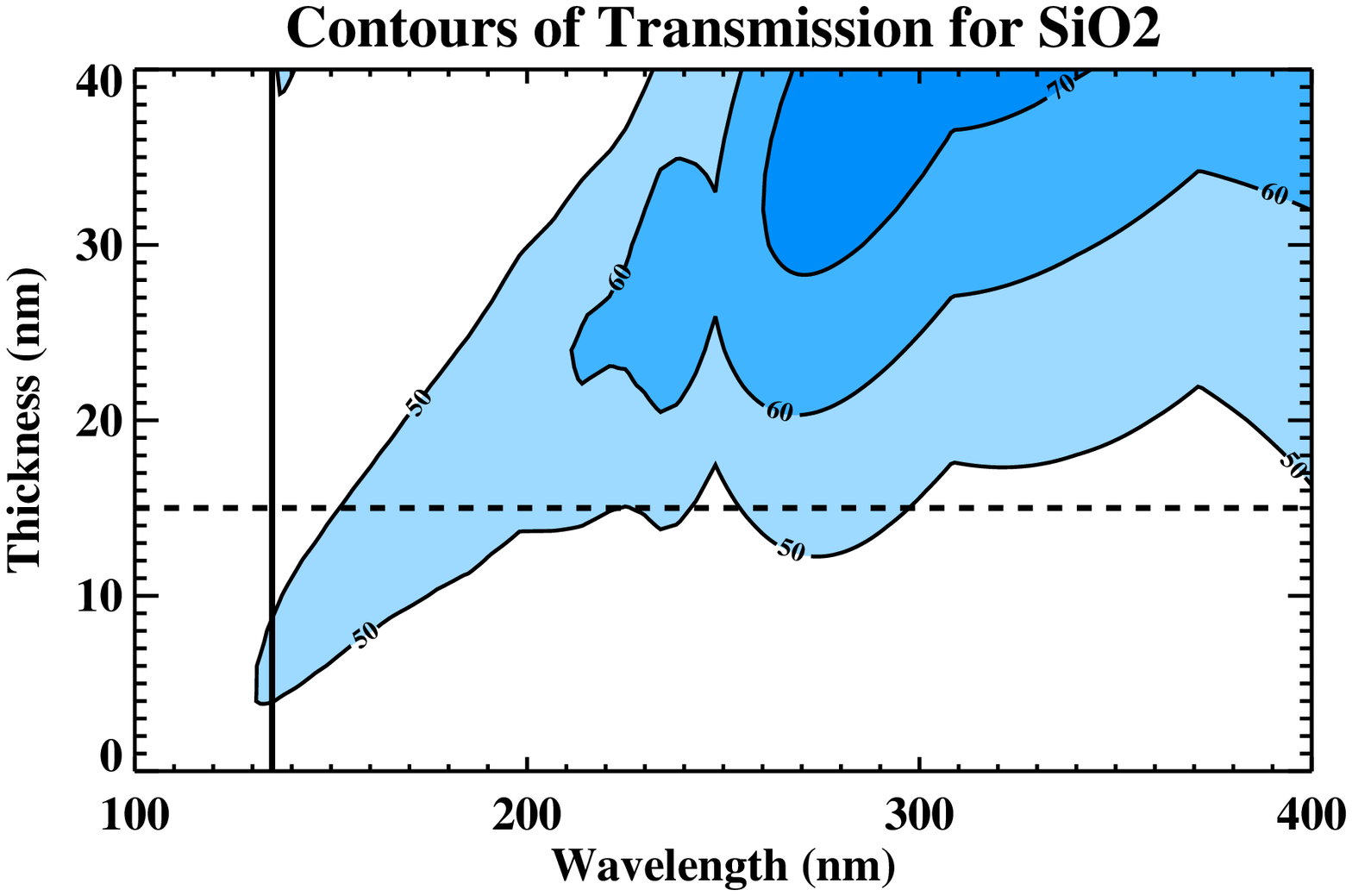}
\caption{Theoretical Transmission for a variety of thicknesses of SiO$_2$.  Contour lines begin at 50 percent transmission and increase in increments of 10 percent.  Horizontal line indicates thickness target.  Vertical line indicates absorption edge.  Absorption increases rapidly as wavelength decreases.}\label{fig:peakSiO2}
\end{figure}

\begin{figure}
\centering
\includegraphics[width=0.45\textwidth,bb= 56 231 576 576]{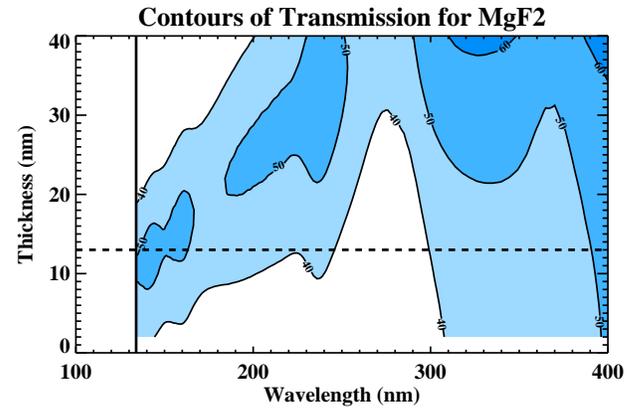}
\caption{Theoretical Transmission for a variety of thicknesses of MgF$_2$.  Contour lines begin at 40 percent transmission and increase in increments of 10 percent.  Horizontal line indicates thickness target.  Vertical line indicates absorption edge.  Absorption increases rapidly as wavelength decreases.}\label{fig:peakMgF2}
\end{figure}

The thickness of the model film was varied until a minimum of losses (reflectance back to the observer plus absorption) was achieved in the target wavelength range.  Contour plots showing potential transmission percentage as a function of wavelength and thickness are shown for each film in Figures \ref{fig:peakHfO2} through \ref{fig:peakMgF2}.  Each plot shows contours of 50-80 percent potential transmission, given a range to thicknesses. A dark vertical line on the plot indicates the absorption edge.  This edge marks the region where absorption begins to increase rapidly as wavelength decreases; generally this fast increase begins when absorption has reached 10-20\%.  Thus anything to the right of the line indicates wavelengths where absorption is not a primary concern.

Thicknesses which provided close to peak transmission while also maintaining a wide range above 50 percent were selected.  The reflectance that corresponded to this thickness was then used as a target during testing.  We sought to make a range of film thickness that were centered around this target.  Typically we tested several films that varried in thickness between 5-10nm above and below the target.  There are several published indices of refraction for the Si substrate, which contributes to some uncertainty in the predicted reflectance.  The indices of refraction for all materials were taken from Palik (\citep{1985P}, \citep{1991P}), except for HfO$_2$, which came from Zukic \citep{1990Z}.  An alternative index of refraction for Si was also consulted, taken from Philipp \citep{1960Philipp}.  The Philipp values differed at the shortest wavelengths (below 150nm) and predicted a lower reflectance than the Palik values.

\begin{table}
\centering
\begin{tabular*}{0.5\textwidth}{@{\extracolsep{\fill}}|c | c c c|}
\hline
Wavelength & & Modeled ideal & Best thickness based on  \\
Range (nm) & Material & thickness (nm) & reflectance test (nm) \\
\hline
240-300 & HfO$_2$ & 22 & 25 \\
200-250 & MgO & 17.5  & 20 \\
180-240 & Al$_2$O$_3$ & 10 & 16 \\
150-180 & SiO$_2$ & 15 & 19 \\
120-150 & MgF$_2$ & 7 & 11 \\
120-150 & MgF$_2$ & 13 & 11 \\
\hline
\end{tabular*}
\caption{Targeted thickness and best actual thickness after reflectance measurements for all films}\label{tab:bestfilm}
\end{table}

All RF Sputtered depositions were made using 15 sccm of argon gas flowing into the deposition chamber kept at a pressure of 25 mTorr during sputtering.  Deposition times varied for each material and between runs.  Deposition rates were calculated based on the measured thickness of a witness sample and the time of deposition.  All targets were conditioned for 3 minutes before sputtering began.  The substrate temperature for all runs was 17$^\circ$ Celsius, except for so called ``hot runs'', which maintained a temperature of 180$^\circ$ Celsius.  In these cases, a 5 minute ``soak'' time was added to ensure the substrate reached equilibrium at deposition temperature.  SiO$_2$, MgO, and Al$_2$O$_3$ were deposited at a power of 480 W (80\% of maximum), while HfO$_2$ was deposited at 252 W (42\% of maximum). 

Thermal evaporation was conducted at pressures of 10$^{-7}$ Torr.  Deposition rates for each run varied between 1 and 4 \AA/sec.  A crystal monitor displayed thickness and an automatic shutter closed when the desired thickness was reached. There was some discrepancy between the monitor thickness and the actual thickness due to a tilt in the substrate holder.  Compensation was made for this discrepancy.

ALD deposition can be performed using either a plasma or a thermal recipe.  The deposition of HfO$_2$ was performed using a plasma recipe nominally at 250$^\circ$ C, while Al$_2$O$_3$ was deposited also using a plasma, but one nominally at 300$^\circ$ C.  Plasma ALD deposition recipes developed here were informed by the work of Goldstein, 2008 \citep{2008Goldstein} for Al$_2$O$_3$ and Lui, 2005 \citep{2005Liu} for HfO$_2$.  

Layers were deposited as close to the selected thickness as could be made, given variations in deposition rate from run to run, and the thickness was verified in an ellipsometer.  Precision of the thickness measurement was 0.1nm. 

\section{Testing}
\label{sec:testing}

\begin{figure}
\centering
\includegraphics[width=0.45\textwidth,bb= 56 231 576 576]{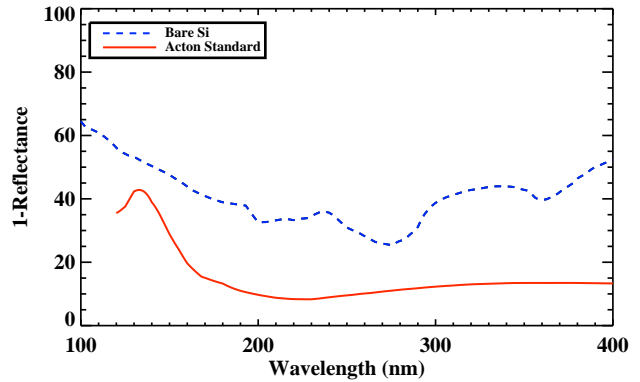}
\caption{1-Reflectance of Acton standard and bare Si.}\label{fig:acton}
\end{figure}

All samples were tested at Columbia University in a reflectance setup.  The samples were placed in a chamber maintained at less than $1 \times 10^{-3}$ torr for the duration of the measurement.  An Acton monochrometer fed by a focused deuterium lamp stepped through wavelengths from 100nm to 300nm in steps of 5nm to 15nm, depending on the testing run.  The light reflecting off of the samples was detected by a Princeton/Acton CCD during exposures of 115 to 180 seconds, again depending on the run.  The resulting images were then dark subtracted, and the counts from the illuminated area extracted.  An Acton H1900-FS-1D standard and an uncoated bare Si wafer were used to calibrate each set of measurements.  Figure \ref{fig:acton} shows the measured (Acton) and predicted (Si) 1-Reflectance percentage of both over the relevant wavelength range.

Films were also sent out for testing by the J.A. Woollam company.  They provided thickness measurements and measures of the optical indices down to 137nm in a vacuum ellipsometer for single samples of HfO$_2$ (both sputtered and ALD), MgO, Al$_2$O$_3$, and MgF$_2$.  The returned values confirmed the expected behavior of all films, except HfO$_2$.  We found the index of refraction below 250nm was much higher than that measured \citep{1990Z} (maximum difference of 30\%).  Values for $k$ were as expected.  We have plans to further test this result.

\subsection{Results}
\label{sec:results}

\begin{figure}
\centering
\includegraphics[width=0.45\textwidth,bb= 56 231 576 576]{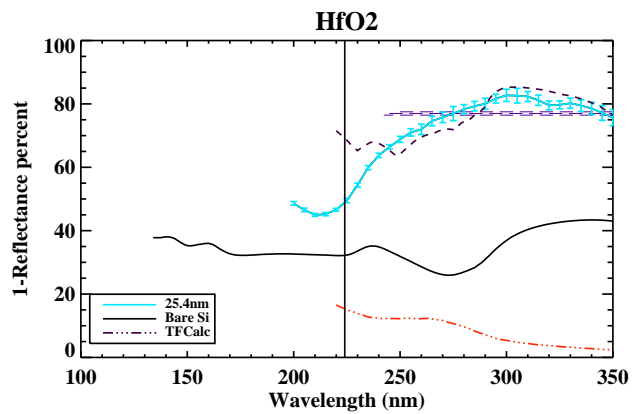}
\caption{1-Reflectance of optimal HfO$_2$ film compared to TFcalc model.  HfO$_2$ has been optimized for 240-300 nm.  The solid vertical line indicates the absorption edge.  The red dot-dashed line shows the corresponding absorption percentage.}\label{fig:HFO2}
\end{figure}

\begin{figure}
\centering
\includegraphics[width=0.45\textwidth,bb= 56 231 576 576]{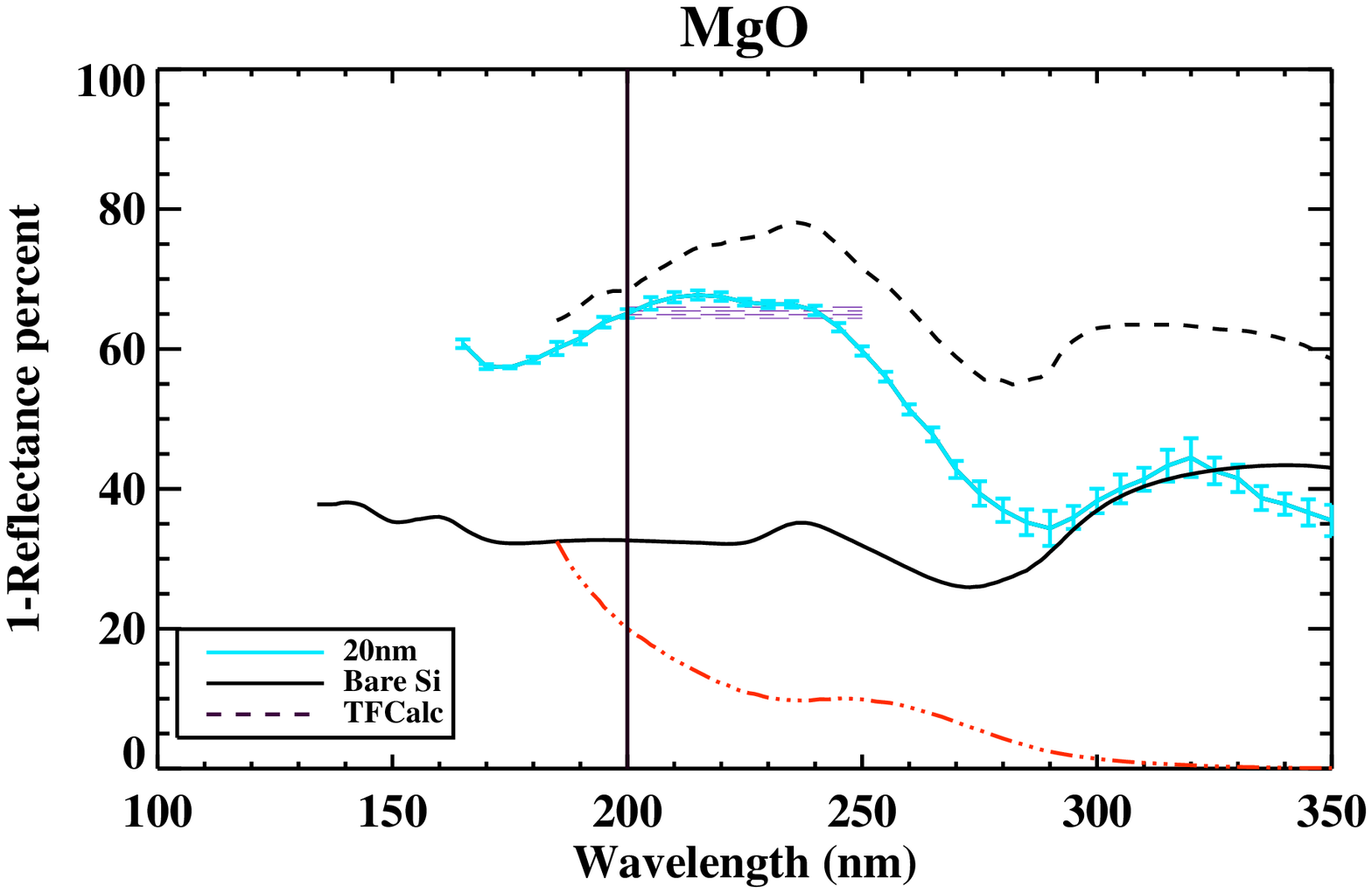}
\caption{1-Reflectance of optimal MgO film compared to TFcalc model.  MgO has been optimized for 200-250nm.  The solid vertical line indicates the absorption edge.  The red dot-dashed line shows the corresponding absorption percentage. }\label{fig:mgo}
\end{figure}

\begin{figure}
\centering
\includegraphics[width=0.45\textwidth,bb= 56 231 576 576]{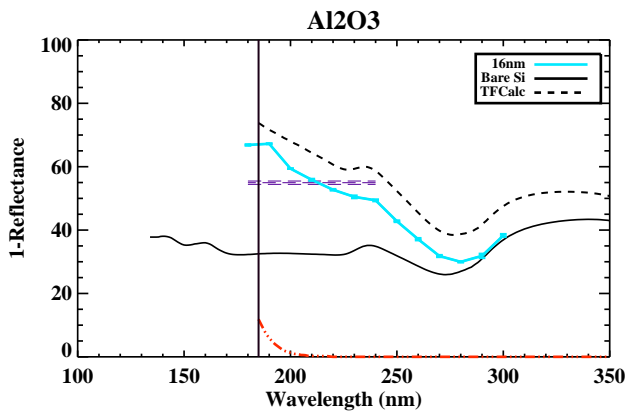}
\caption{1-Reflectance of optimal Al$_2$O$_3$ film compared to TFcalc model.  Al$_2$O$_3$ has been optimized for 180-240 nm.  The solid vertical line indicates the absorption edge.  The red dot-dashed line shows the corresponding absorption percentage. }\label{fig:Al2O3}
\end{figure}

\begin{figure}
\centering
\includegraphics[width=0.45\textwidth,bb= 56 231 576 576]{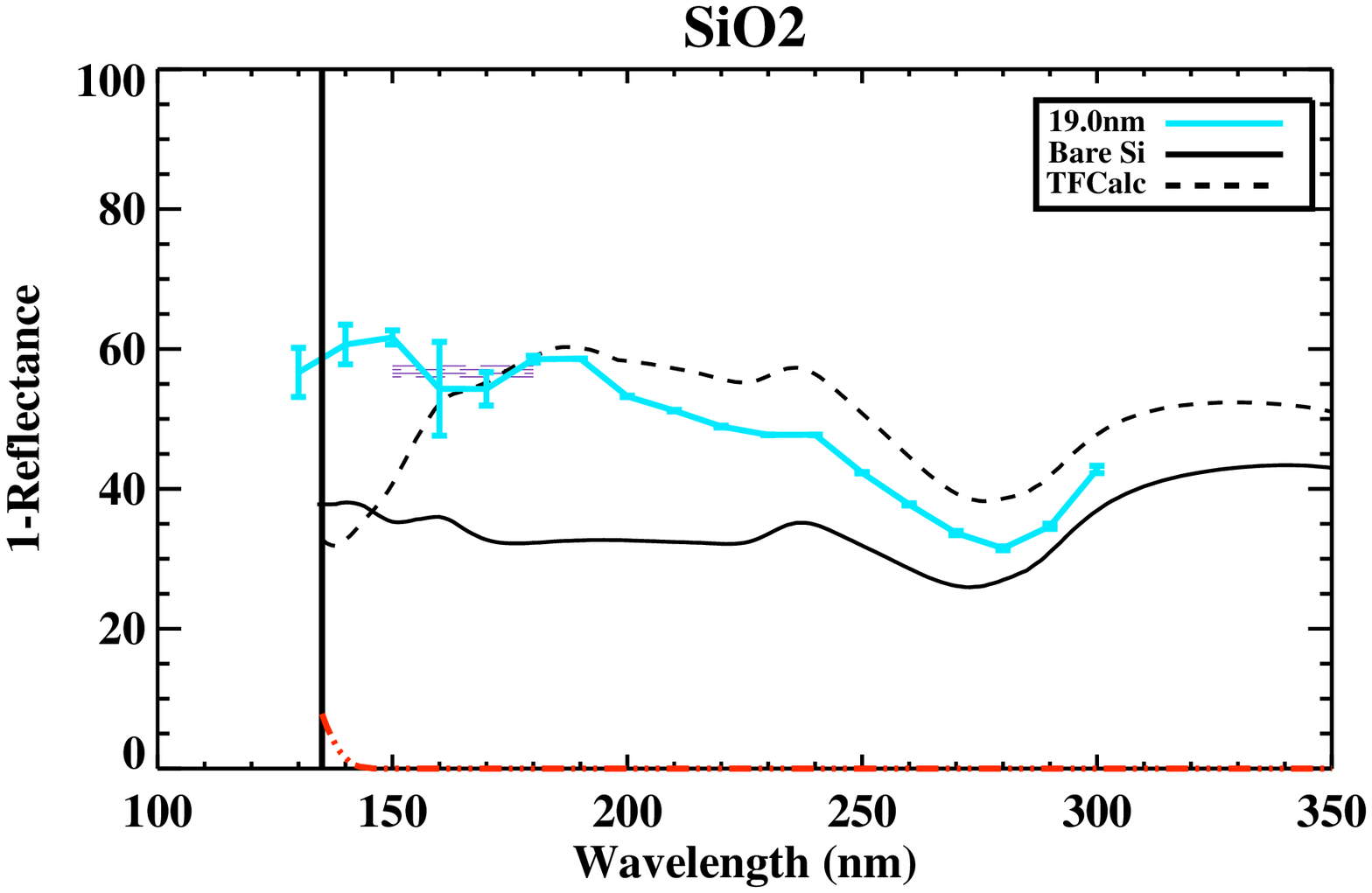}
\caption{1-Reflectance of optimal SiO$_2$ film compared to TFcalc model.  SiO$_2$ has been optimized for 150-180 nm.  The solid vertical line indicates the absorption edge.  The red dot-dashed line shows the corresponding absorption percentage.}\label{fig:Sio2}
\end{figure}

\begin{figure}
\centering
\includegraphics[width=0.45\textwidth,bb= 56 231 576 576]{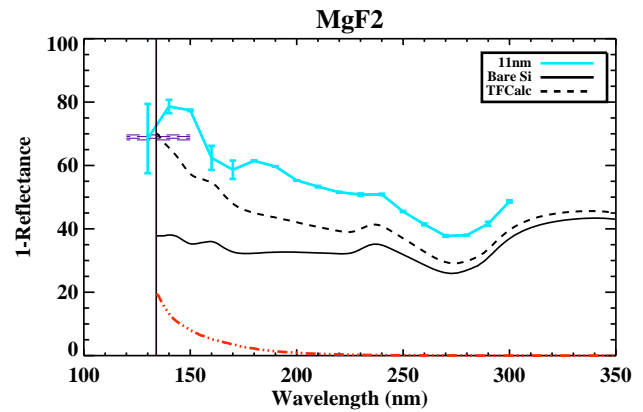}
\caption{1-Reflectance of optimal MgF$_2$ film compared to TFcalc model.  MgF$_2$ has been optimized for 120-150 nm.  The solid vertical line indicates the absorption edge.  The red dot-dashed line shows the corresponding absorption percentage. }\label{fig:mgf2}
\end{figure}

After comparing the reflectance of different thicknesses of the same material, a best fit layer thickness was selected.  This best fit for each material (listed in Table \ref{tab:bestfilm}) depended on how well the layer minimized reflectance over the wavelength range in question, as well as an adherence to the predicted behavior of the reflectance curve.  We were able to achieve predicted behavior in almost all of the wavelength ranges, with the exception of SiO$_2$.  The results of the initial tests are presented in Figures \ref{fig:HFO2} through \ref{fig:mgf2}.  Figures are plotted in terms of 1-Reflectance, which over the wavelengths of interest is a proxy for Transmittance.  A vertical solid line passes through the figure at the point at which absorption reaches 10\%.  At wavelengths lower than the line, absorption grows rapidly and we assume the bulk of diminished reflectance is due to absorption and not transmission.  For the sake of clarity, we present 1-Reflectance for only our best fit thickness.

Further tests were done to ensure reliability.  Two to three films were made for each material with the same deposition process as the best fit film.  Two things were tested here: the repeatability of the deposition procedure and consistency of performance from sample to sample.  Deposition of all have proven to be quite repeatable (thicknesses vary by ~1 nm from target) and are especially repeatable for ALD depositions.  The values for reflectance are also consistent across samples of the same material.

Materials deposited using ALD, as opposed to sputtering, appear to produce better films.  The thickness of the film is easier to control and the quality of the layer (as judged by how close the index of refraction at 500nm is to the bulk index) was consistently higher.  Furthermore, on tests with live delta-doped devices, sputtering caused a variety of problems.  We believe interactions of the film material with the surface Si created highly absorptive silicates.  As such, our later work has used ALD exclusively for HfO$_2$ and Al$_2$O$_3$.  We plan to transition to ALD for MgO and SiO$_2$ once the necessary precursors have been obtained.  A forthcoming paper will discuss in more detail the comparisons between films made by ALD and sputtering (Greer, 2011 (submitted)).

\begin{figure}
\centering
\includegraphics[width=0.45\textwidth,bb= 56 231 576 576]{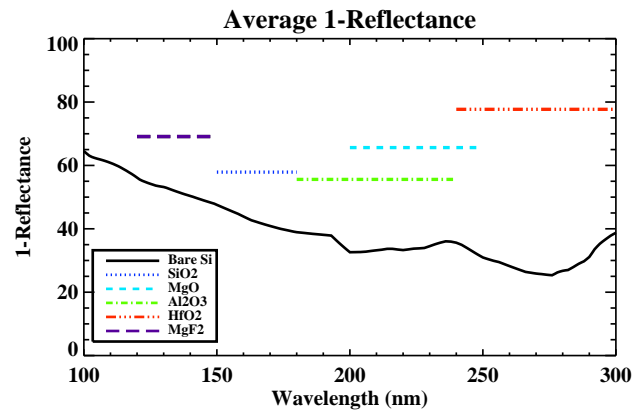}
\caption{Average 1-Reflectance of all films.  Solid bars indicate the average of 1-Reflectance over the wavelength range of interest for each material.}\label{fig:ave}
\end{figure}

AR coating films have shown good stability thus far and long term stability tests are underway, to be reported separately.  Tests remain to determine the feasibility of applying different non-overlapping layers to the same CCD. Figure \ref{fig:ave} depicts the overall outlook for multiple coatings on a single CCD.  Each film is shown over the wavelength range of interest.  The solid bar is an average of the reflectance values over that range.  This provides an encouraging outlook for a future detector.

\section{Testing on Live Devices}
\label{sec:devices}

Our team has applied the films discussed above to thinned and delta-doped CCDs.  Work on this is ongoing and another publication describes our results in detail, but it merits a short mention here.

The procedure for thinning and delta-doping, as described in Hoenk, \citep{2009Hoenk}, was carried out in the MicroDevices Lab at JPL on both Cassini CCDs and Electron Multiplying CCDs (EMCCDs) \citep{2001Mackay}.  The standard n-channel CCDs used have 1024 square arrays and 12 micron pixels \citep{2004Porco}.  The EMCCDs are 1024 by 512 rectangular arrays, with 16 micron pixels.  EMCCDs are an advancement in CCD technology which enables low-noise photon-counting. These were first developed at e2v (formally Marconi), and sport a high internal gain before readout, which allows sub electron read noise levels.  The advantages for registering single photon events are obvious \citep{2004Gach}, and when combined with a high QE from a thinned, delta-doped, and AR coated CCD, we expect to dramatically improve the minimum signal detector threshold.  

For each test, a single film is applied to a thinned and delta-doped device, using a simple shadow mask to provide a coated and uncoated half.  Devices were then wire-bonded and packaged for testing.  A measurement set-up for finding absolute QEs was developed using JPL's Vacuum UV Characterization setup (a detailed description of characterization system and methods of measurements are described in a separate paper (Jacquot, Rev Sci Instrum, submitted)) and used for measuring the effectiveness of each film type.  The testing chamber steps through wavelengths from Lyman-$\alpha$ into the infrared, illuminating the CCD.  A NIST-calibrated Si photodiode provides a standard at each wavelength.  A region of interest was selected on both the coated and uncoated sides, and QE was calculated from the counts in each region after correcting for bias levels.  The conversion gain ($E/DN$) is determined by a photon transfer curve.  The uncoated QE should follow the reflection limit of silicon, and so provides an excellent check that the device is functioning properly.

Films made with RF sputtering failed to increase QE, and in some cases caused worse performance.  ALD and Thermal Evaporated films all performed well.  Using parameters from the literature we have achieved qualitative agreement between our measured QE and the expected value.  In films where we characterized the indices, we achieved more close agreement between modeling and experiments.  More work is underway to complete the analysis.  Nevertheless, we have been able to achieve greater than 50\% QE over the range of interest in the four bands described previously.  More detailed description of the behavior of the coatings and plots of QE of the films on live devices will be discussed in a future paper.

\section{Future Applications}
\label{sec:future}

\subsection{Applications in astronomy}

Astronomical uses for these new devices will no doubt go beyond the few we have imagined here, as more work is done on these and other materials.  The most immediate use we envision would be on a space, balloon, or rocket borne experiment in which the device is fed by an Integral Field Unit (IFU) and will cover a wavelength range of 120-300nm.  This would allow imaging and spectroscopy beyond the frontiers created by GALEX and HST, in a wider range of wavelengths and at much lower integration times for similar information.  The use of an EMCCD would allow observations of the faintest regions of the intergalactic medium, and could provide an important view into the history of the Universe.

Beyond an IFU covering a wide wavelength range, these coatings can be tuned to any variety of specific medium bands.  An AR-coating is already in use on WFPC3, providing QE in the NUV and optical \citep{2010B}.  The coverage of a whole device in a thick layer of Al2O3 (23nm), for instance, would create an imager optimized for 200-250nm.  HfO$_2$, already well known as an AR-coating in the optical, could similarly be useful in a near UV and optical imager.  MgO and MgF$_2$, along with other as of yet untested materials, may also prove useful in specific instances.  We plan to test more films and thicknesses on live devices to get a better idea of the potential range of uses.

\subsection{Multi-layer and Graded Coatings}

\begin{figure}
\centering
\includegraphics[width=0.45\textwidth,bb= 56 231 576 576]{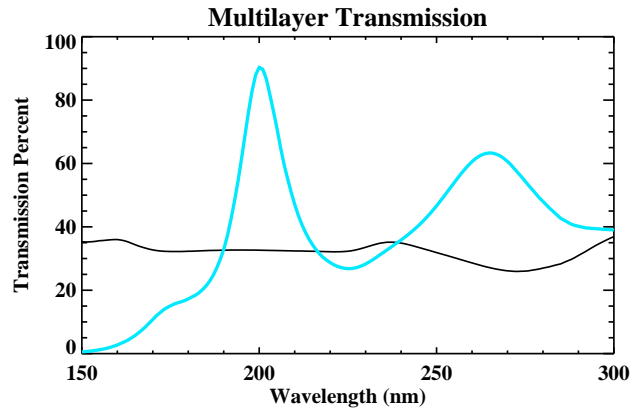}
\caption{Theoretical Transmission for a 5 layer stack of LaF$_3$ and Al$_2$O$_3$. This particular multilayer is optimized to maximize transmission around 200nm.  1-Reflectance of Si is shown as solid black line.}\label{fig:Multilayer}
\end{figure}

Thus far we have only discussed the effects of single films.  One may also create narrow-band filters in the UV using these coatings.  As is well known, multi-layer coatings with alternating layers of high and low index material can provide even better transmission than a single film AR coating (\citep{1982F}, \citep{1976H}, \citep{1972S}).  A highly tuned multi-layer, while being less broadly applicable, can provide a bandpass of high transmission over a short wavelength range.  A multi-layer film using alternating layers of LaF$_3$ and MgF$_2$ is one such easily created design.  These multi-layers could be optimized to create a bandpass around 200nm, taking advantage of an atmospheric window best exploited by balloon experiments \citep{1980Huffman}. This bandpass would have higher transmission than a single layer of MgO or Al$_2$O$_3$, but would be much narrower in wavelength.  This particular example is ideal for looking at red-shifted Lyman-$\alpha$ and certain red-shifted metal lines while taking advantage of the very low sky background found in the UV \citep{2008Tuttle}, but coatings can be optimized for nearly any desired wavelength and observational purpose.  Figure \ref{fig:Multilayer} shows an example of the high, but peaked, transmission that can be achieved with a multilayer film.  LaF$_3$ and Al$_2$O$_3$ are used in sequence and with different thicknesses in each layer.  There are 5 layers all together, starting and ending with Al$_2$O$_3$.   

In our work so far, we have tried to achieve high QE across a wide wavelength range by using several different films, each with a specific thickness.  Each film, while not being a true broadband coating, still provides coverage over many tens of nanometers.  Taking this method one step further, one could create a series of many very sharply peaked narrow-band films, and similarly tile them across a device.  Each would provide higher QE than the films described in previous sections, but only over a range of a few nanometers.  This creates technological challenges in applying these films over a small area and in close proximity.  One alternative to this would be to create a graded or ramped coating such that the thickness changes quickly enough to provide high QE without changing film materials.  A wedge shaped film (thin at low wavelengths and thicker at high) would eliminate the need to change from one material type to another, and provide a way to reach a consistently high QE across the whole band instead of just one peak.  This ramp would also prevent the creation of low QE 'seams', where two films meet.  

\begin{figure}
\centering
\includegraphics[width=0.45\textwidth,bb= 56 231 576 576]{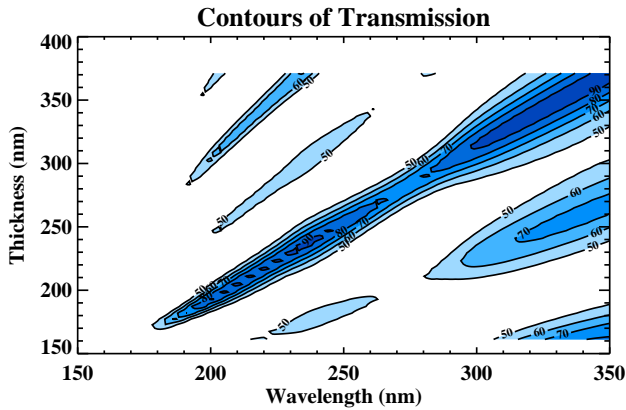}
\caption{Theoretical Transmission for a variety of total thicknesses of stacks of LaF$_3$ and Al$_3$O$_3$.  The 5-layer stacks are varied to achieve high QE in a narrow range.  A film that was graded to follow the highest contours could achieve between 75 and 95 percent transmission across a large wavelength range.}\label{fig:Multilayer_contours}
\end{figure}

One example of a potential ramp is found in Figure \ref{fig:Multilayer_contours}.  The modeled coating is 5 layers of alternating Al$_2$O$_3$ and LaF$_3$.  While the ratios between the layers remain roughly static, the overall thickness of the stack increases as one optimizes for higher wavelengths.  By starting at one edge of the device with a coating optimized for 180nm and increasing all thicknesses until the coating is optimized for 350nm, for example, one can create a device with high QE across all wavelengths.  The material combinations for this type of design are dependent on what ranges one hopes to target, and are somewhat constrained by deposition techniques, but on the whole this represents a way to achieve numerous goals in high QE coatings.

\subsection{Red Rejection}

A common difficulty in UV spectroscopy is adequate red rejection.  All of the AR coatings we have discussed above provide a boost in transmission in the UV, but at longer wavelengths has very little effect.  Thus at the red end, the CCD should behave as an uncoated delta-doped device and be very responsive.  Any optical assembly which seeks to use these devices will need to ensure that stray light of other wavelengths does not reach the CCD.  The high QE over a wide range of wavelengths longer than the UV means this is a particularly important issue to note.  Filters, combined with a careful selection of reflective gratings, should minimize red leak, although characterization of the total leak into the device on the final optical assembly will be essential.  We may also explore filters that simultaneously reflect undesired optical bands and transmit effectively in the UV.  

\subsection{Applications in other fields}

While our expertise remains in astronomical and scientific applications, the uses of a high quality imager over the entire UV wavelength range will be considerable.  The devices are relatively easy to manufacture on an industrial scale, especially when compared with MCPs.  One immediate application is in small scale modular fixed spectrographs.  A variety of these are made by Ocean Optics (USB2000 series), StellarNet (EPP2000), and other companies.  Judicious use of AR-coatings for these modular spectrographs will immediately and easily improve performance.  VUV spectrographs from Acton/PI and McPherson and VUV ellipsometers from J.A.Woollam, among others, can also see similar improvements.  Instruments which already utilize spectrometers to identify chemicals in the fields of medical imaging, defense, materials analysis, and chemical testing can also take advantage of the coatings presented here.

Any potential uses one might imagine are immediately bolstered by the increased QE that these devices will provide.  From an astronomical perspective, a boost in QE from delta-doping and AR coating, combined with a decrease in the noise level from an EMCCD, will result in two advantages.  The first is a lowering of detection limits, allowing deeper images of fainter objects.  The second is a cost savings on future missions.  A smaller primary mirror could be used while still achieving great scientific gains and lowering the overall costs of getting a mission into the upper atmosphere or orbit.  Recent developments in creating low noise CMOS detectors mean our AR coatings may also be useful in this brand of UV detector development \citep{2010F}.  In other fields, using relatively easy to manufacture silicon detectors in place of MCPs or other more construction-demanding devices will also provide a savings in cost, design, and ease of use.

\section*{Acknowledgements}

\noindent The research described here was funded in part by a NASA Space Grant.

The research described in this paper was carried out in part at the Jet Propulsion Laboratory, California Institute of Technology, under a contract with the National Aeronautics and Space Administration.

The research here was supported in part by internal funding from Columbia University.

\bibliography{arlib.bib}

\end{document}